\documentclass[aip,jcp,graphicx,preprint]{revtex4-1}
\usepackage{graphicx}
\usepackage{amsmath,amssymb}
\usepackage{mhchem}
\usepackage{color}
\usepackage[colorlinks={true}]{hyperref} 
\hypersetup{citecolor={blue}, filecolor={blue}, linkcolor={blue}, urlcolor={blue}}
\usepackage{dcolumn}
\usepackage{bm}

\begin{document}


\title{Dissociation dynamics of 3- and 4-nitrotoluene radical cations: Coherently driven C$-$NO$_2$ bond homolysis} 



\author{Derrick Ampadu Boateng}
\affiliation{Department of Chemistry, Virginia Commonwealth University, Richmond, VA 23284, USA}
\author{Gennady L. Gutsev}
\affiliation{Department of Physics, Florida A\&M University, Tallahassee, FL 32307, USA}
\author{Puru Jena}
\affiliation{Department of Physics, Virginia Commonwealth University, Richmond, VA 23284, USA}
\author{Katharine Moore Tibbetts}
\email{kmtibbetts@vcu.edu}
\affiliation{Department of Chemistry, Virginia Commonwealth University, Richmond, VA 23284, USA}

\date{\today}

\begin{abstract}
Monosubstituted nitrotoluenes serve as important model compounds for nitroaromatic energetic molecules such as trinitrotoluene. This work investigates the ultrafast nuclear dynamics of 3- and 4-nitrotoluene radical cations using femtosecond pump-probe measurements and the results of density functional theory calculations. Strong-field adiabatic ionization of 3- and 4-nitrotoluene using 1500 nm, 18 fs pulses produces radical cations in the ground electronic state with distinct coherent vibrational excitations. In both nitrotoluene isomers, a one-photon excitation with the probe pulse results in \ce{NO2} loss to form \ce{C7H7+}, which exhibits out-of-phase oscillations in yield with the parent molecular ion. The oscillations in 4-nitrotoluene with a period of 470 fs are attributed to the torsional motion of the \ce{NO2} group based on theoretical results showing that the dominant relaxation pathway in 4-nitrotoluene radical cation involves the rotation of the \ce{NO2} group away from the planar geometry. The distinctly faster oscillation period of 216 fs in 3-nitrotoluene is attributed to an in-plane bending motion of the \ce{NO2} and \ce{CH3} moieties based on analysis of the normal modes. These results demonstrate that coherent nuclear motions determine the probability of \ce{C-NO2} homolysis in the nitrotoluene radical cations upon optical excitation within several hundred femtoseconds of the initial ionization event.
\end{abstract}


\maketitle 

\section{Introduction}\label{intro}
Understanding the detonation mechanisms of energetic materials used in military \cite{Sikder2004} and mining \cite{Wharton2000} operations has been an active area of research for decades. While extensive studies have unraveled mechanisms of stored energy released in bulk energetic materials, \cite{Field1982,Tarver1997,Dremin2000,Ramaswamy2001,Cohen2007} initial dissociation mechanisms of isolated energetic molecules still constitute an active area of investigation.\cite{Jeilani2015,Zeng2016,Yuan2015,Yuan2016,Yuan2017} In energetic materials, excited electronic states and molecular ions are thought to drive initial energy release processes based on the observation of tribological luminescence.\cite{Zink1973,Lin1980} Thus, understanding relaxation and dissociation processes from excited states and ions of isolated energetic molecules is needed to fully understand initial excitation events in energetic materials, which may facilitate longstanding goals such as developing photoactive high explosives that can be initiated by lasers.\cite{Bowden2007} 

Because ultrafast events associated with the initial dissociation pathways of energetic molecules typically occur within a few picoseconds of the initial excitation,\cite{Dlott2003} time-resolved pump-probe methods originally developed by Zewail\cite{Zewail1988} are needed to investigate their dynamics. Pump-probe studies have revealed the dynamics of fast evolving events in energetic molecules including HMX and RDX,\cite{Greenfield2006} nitramines,\cite{Guo2005,Guo2007,Guo2011} nitromethane,\cite{Guo2009} and furazan.\cite{Guo2008} In all of these molecules, the ionized fragment \ce{NO+} was formed from the electronically excited parent molecules within the pulse duration of 180 fs. In other studies, dissociation of \ce{NO2} from nitromethane was found to occur within 81 fs\cite{Nelson2016} and the transient parent cations of nitrotoluenes formed from electronically excited neutrals were found to have lifetimes of $50-70$ fs.\cite{Wang2010} While these studies and others attest to ultrafast timescales leading to decomposition of energetic molecules from their neutral excited states, less is known about the dissociation dynamics of their radical cations. It is thought that the radical cations of energetic molecules dissociate via low-lying electronic states based on photoelectron-photoion coincidence measurements of nitromethane\cite{Ogden1983} and nitrobenzene,\cite{Panczel1984} as well as shaped 800 nm femtosecond laser pulse excitation of 4-nitrotoluene.\cite{Lozovoy2008} Recent theoretical studies have also identified ground state dissociation pathways in the radical cations of 1-nitropropane\cite{Tsyshevsky2014} and trinitrotoluene (TNT).\cite{NguyenVan2015} However, the timescales of these dissociation processes remain unknown. 

One of the most important families of energetic molecules is the nitrotoluenes, with TNT the most widely investigated due to its practical uses.\cite{Wharton2000,Sulzer2008,NguyenVan2015,Cohen2007,McEnnis2007,Mullen2009,Furman2016,Weickhardt2002} While multiple rearrangement reactions can occur in excited TNT and other nitrotoluenes, the homolysis of one or more \ce{NO2} groups has been found to drive initial detonation processes based on its thermodynamic favorability.\cite{Cohen2007} As model compounds for TNT, the dissociation of mononitrotoluenes using nanosecond and femtosecond laser pulses of various wavelengths has been widely investigated.\cite{Kosmidis1994,Kosmidis1997,Tasker2002,Weickhardt2002,Lozovoy2008} Ionization with nanosecond laser pulses results in only small fragments observed in the mass spectra owing to the short-lived neutral electronic excited states of these compounds. \cite{Kosmidis1994,Kosmidis1997} In contrast, femtosecond laser ionization results in less fragmentation because the ion is formed via multiphoton ladder climbing before dissociation from the neutral excited state.\cite{Kosmidis1997,Tasker2002,Lozovoy2008} However, even under multiphoton ionization conditions, the parent nitrotoluene ion is rarely the dominant peak in the mass spectrum,\cite{Kosmidis1997,Weickhardt2002,Tasker2002,Lozovoy2008} which renders study of its dissociation via pump-probe methods difficult.

The limited formation of the parent molecular ion in nitrotoluenes and other polyatomic molecules arises because strong field multiphoton ionization is a nonadiabatic process that can populate multiple excited states in the cation.\cite{Lezius2001,Lezius2002} In contrast, strong field ionization via electron tunneling can form predominantly ground state molecular ions because a limited amount of energy is injected into the remaining ion during electron detachment. \cite{Lezius2001,Lezius2002} Tunnel ionization in atoms was first explained by Keldysh, where the transition from nonadiabatic to adiabatic ionization was described by the Keldysh adiabaticity parameter $\gamma$ given by the ratio of the laser frequency $\omega_0$ to the electron tunneling frequency $\omega_t$,\cite{Keldysh1965}
\begin{equation}
\gamma=\frac{\omega_0}{\omega_t}=\omega_0\frac{\sqrt{2\Delta m_e}}{eE_0},\label{keldysh}
\end{equation}
where $\Delta$ is the ionization potential, $m_e$ and $e$ are the electron mass and charge, respectively, and $E_0$ is the laser electric field strength. The case $\gamma>>1$ corresponds to a high laser frequency that inhibits electron tunneling through the electrostatic potential barrier before the electric field switches signs. This situation results in nonadiabatic multiphoton ionization as the electron continues to absorb energy over multiple cycles of the laser pulse.\cite{Lezius2001,Lezius2002} Alternatively, the probability for electron tunneling is greatly increased when $\gamma<<1$, resulting in adiabatic ionization that imparts little energy to the remaining electrons.\cite{Lezius2001,Lezius2002} Even though the Keldysh theory was proposed for atoms, it has also been found viable for polyatomic molecules with recent experiments of strong field ionization using near-infrared excitation wavelengths (e.g., $1150 - 1600$ nm) where $\gamma<<1$.\cite{Lezius2001,Lezius2002,Yatsuhashi2005,Murakami2005,Tanaka2009,Bohinski2013a,Bohinski2013b,Bohinski2014a,Tibbetts2014,Bohinski2014b,Tibbetts2015,Munkerup2017,AmpaduBoateng2018} For example, ionization of decatetraene\cite{Lezius2002} and anthracene\cite{Murakami2005} with 800 nm excitation resulted in extensive fragmentation as compared to using 1400 nm excitation, where the enhanced fragmentation was attributed to cationic resonances and nonadiabatic ionization dynamics. Similar results were obtained for acetophenone excited with wavelengths between 800 nm and 1434 nm, where significantly less dissociation to small fragments was observed for long wavelengths.\cite{Bohinski2013b} Recent pump-probe studies have demonstrated improved preparation of coherent vibrational wavepackets in ground state molecular ions when using $1200-1500$ nm instead of 800 nm pulses for ionization. \cite{Bohinski2014b,Tibbetts2015,AmpaduBoateng2018}

This work will present pump-probe measurements on the radical cations of the isomeric compounds 3-nitrotoluene and 4-nitrotoluene (3-NT and 4-NT, respectively). The high yield of parent molecular ion when using 1500 nm pulses for ionization enables observation of coherent nuclear dynamics in the radical cations of both 3-NT and 4-NT. The ion yields in the two isomers exhibit distinct oscillatory dynamics, indicating coherent excitation of distinct normal modes. Interpretation of the experimental results is supported by a series of density functional theory (DFT) calculations of the optimized geometries, relaxation pathways, and vibrational frequencies in 3-NT and 4-NT radical cations. The remainder of this work is structured as follows: Sections \ref{expt} and \ref{theory} describe the experimental and computational methods, respectively. Sections \ref{results} and \ref{theoryres} present the experimental and theoretical results. Section \ref{disc} presents an interpretation of the observed dynamics, and Section \ref{con} presents concluding remarks.

\section {Experimental methods}\label{expt}
\begin{figure}[htbp]
\renewcommand{\baselinestretch}{1}
\begin{center}
\includegraphics[width=8.5cm]{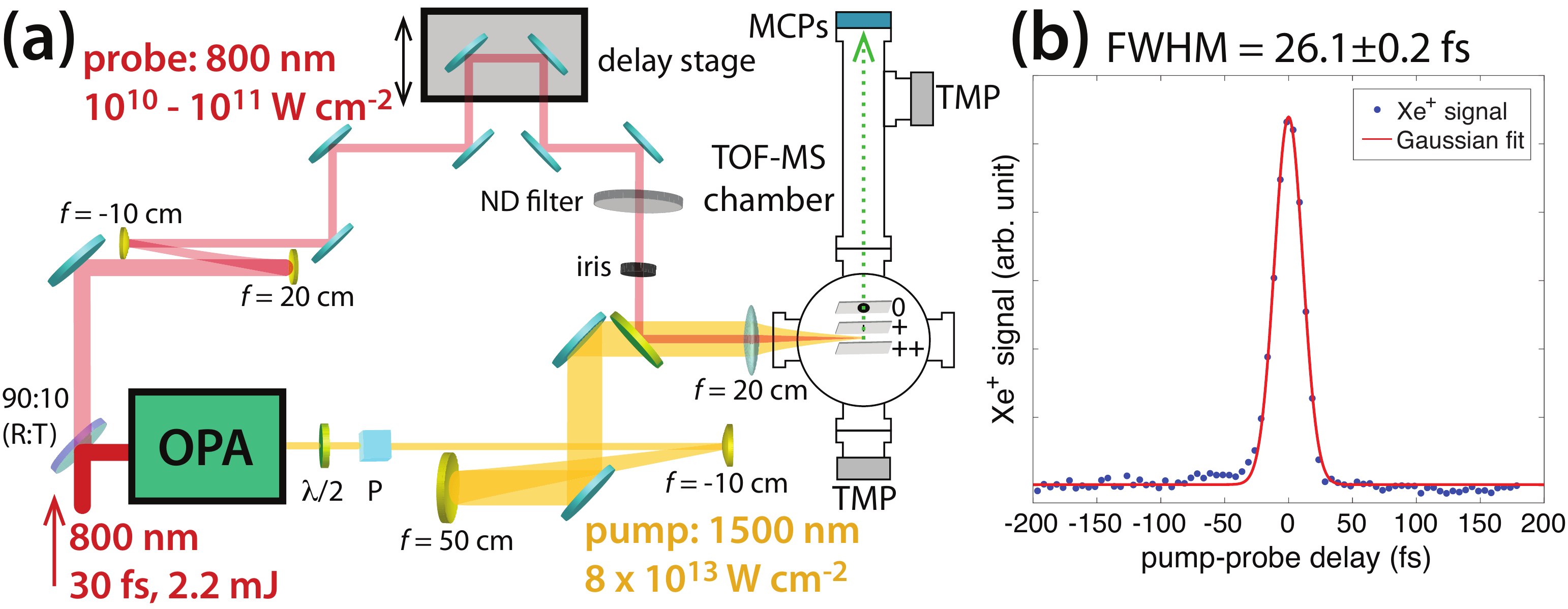}
\end{center}
\caption{\label{setup} (a) Experimental setup showing pump and probe beamlines. (b) Cross-correlation measurements of Xe$^+$ using the $1500$ nm pump and $800$ nm probe. The cross-correlation FWHM of $26.1$ fs obtained by fitting the experimental data (blue dots) to a Gaussian function (red) is consistent with the pulse durations measured by FROG.}
\end{figure}

The pump and probe pulses are generated from a Ti:Sapphire regenerative amplifier (Astrella, Coherent, Inc.) producing 30 fs, 800 nm, 5 mJ pulses. 2.2 mJ of the laser output is split with a 90:10 (r:t) beamsplitter, with 1.9 mJ used to pump an optical parametric amplifier (OPA, TOPAS Prime) to produce $1500$ nm, 18 fs, $300$ $\mu$J pulses that are used as the pump. The pump pulse energy is attenuated with a $\lambda/2$ waveplate and polarizer and expanded using two spherical gold mirrors with $f = -10$ cm and $f = 50$ cm to increase the beam diameter (measured with the knife-edge method) from $4.5$ mm to $22$ mm (Figure \ref{setup}(a), yellow beamline). This beam expansion results in a smaller focal beam waist, and thus higher intensity. The remaining 200 $\mu$J of 800 nm acts as the probe pulse and is down-collimated using two spherical gold mirrors with $f = 20$ cm and $f= -10$ cm to reduce the beam diameter from $11.6$ mm to $5.8$ mm. The probe beam is then directed to a retro-reflector (PLX, Inc.) placed on a motorized translation stage (ThorLabs, Inc.), attenuated with a variable density filter, and passed through an iris to isolate the most intense portion of the beam (Figure \ref{setup}(a), red beamline). Pump and probe beams are recombined on a dichroic mirror and focused with an $f=20$ cm fused silica biconvex lens. The durations of the pump and probe pulses were measured with a home-built Frequency Resolved Optical Gating (FROG)\cite{Kane1993} setup to be 18 fs and 25 fs, respectively (Supplemental Material, Figure S1).

The focused laser pulses are introduced into an ultrahigh vacuum chamber (base pressure $2\times10^{-9}$ torr) coupled to a linear time-of-flight mass spectrometer (TOF-MS, Jordan, Inc.) described in our earlier work.\cite{Gutsev2017} The focus of the laser beam is centered between the charged repeller ($+4180$ V) and extraction ($+3910$ V) plates, where a $1/16$'' diameter stainless steel tube introduces the sample as an effusive molecular beam approximately $1$ cm from the from the laser focus. The extraction plate has a $0.5$ mm slit orthogonal to the laser propagation and TOF-MS axes, which serves as a filter to allow only ions produced at the central focal volume of the laser beam where the intensity is the highest to enter the flight tube.\cite{Hankin2001} Samples of 3-NT and 4-NT (Sigma Aldrich, Inc.) are used without further purification and introduced directly into the chamber. Due to the low vapor pressures of the nitrotoluenes, the sample holders are heated with resistive heating tape to produce a pressure of $10^{-7}$ torr at the Z-gap microchannel plate (MCP) detector. Mass spectra are recorded with a 1 GHz digital oscilloscope at a sampling rate of 20 giga samples per second (GS/s) (LeCory WaveRunner $610$Zi). All mass spectra are averaged over $10,000$ laser shots. 

To establish the time-resolution of our experimental setup, Xe gas was introduced to the chamber and cross-correlation measurements using 1500 nm pump and 800 nm probe pulses with intensities of $6\times10^{13}$ W cm$^{-2}$ and $1\times10^{13}$ W cm$^{-2}$, respectively (Figure \ref{setup}(b)). Fitting the Xe$^{+}$ signal as a function of pump-probe delay to a Gaussian function produced a FWHM of $26.1\pm0.2$ fs, consistent with the FROG measurements. The absolute intensity of the pump pulse was calibrated by measuring the sum of Xe$^{n+}$ ion intensities as a function of pulse energy, which were fit to tunneling ionization rates of a rare gas according to the well-established procedure.\cite{Hankin2001} The absolute intensity of the probe pulse was obtained by measurement of the focused beam waist using a CMOS camera (ThorLabs, Inc). The beam waist and Rayleigh range were determined to be $29.4$ $\mu$m and $16$ mm, respectively (Supplemental Information, Figure S2). With the probe pulse energies of $1-11$ $\mu$J and duration of 25 fs, the peak intensities were calculated to be in the range of $2\times10^{10}-2\times10^{11}$ W cm$^{-2}$.  

\section{Computational methods}\label{theory}
Our computations are performed using the widely used B3LYP method,\cite{Becke1993,Stephens1994} as implemented in Gaussian 09 suite of programs.\cite{Gaussian09}  We choose a balanced split-valence Def2-TZVPP [($11s6p2d1f$)/$5s3p2d1f$] basis  of triple-$\zeta$ quality. The convergence threshold for total energy was set to $10^{-8}$ eV and the force threshold was set to 10$^{-3}$ eV/\AA. Each geometric optimization was followed by harmonic frequency computations in order to confirm the stationary character of the state obtained. In order to test the accuracy of our computational approach, we have optimized the anionic states in addition to the neutral and cationic states of both 3-NT and 4-NT.

The ground state of the neutral 4-NT molecule was found to be lower in total energy than the ground state of neutral 3-NT by 0.018 eV. This difference comes from only electronic total energies since the zero-point vibrational energies match each other within 0.001 eV. The ionization energies computed as the difference in total energies of the cation and its neutral parent at the equilibrium geometry of the neutral are rather close to each other; namely, 9.60 eV for 4-NT and 9.48 eV for 3-NT, which compare well with the experimental values of 9.54 eV\cite{Zhang2012} and 9.48 eV,\cite{Kobayashi1974} respectively.  Our computed electron affinities of the para- and meta-isomers of 1.04 eV and 1.09 eV practically match the experimental values of $0.932 \pm0.087$ eV\cite{Huh1999} and $0.99 \pm0.10$ eV,\cite{Chowdhury1986} respectively, within the experimental uncertainty bars. In view of close agreement of our computed values with experiment, one can expect the B3LYP/Def2-TZVPP approach to be accurate in the same extent when computing other properties of the nitrotoluene isomers.

\section{Experimental results}\label{results}

\subsection{Time-resolved mass spectra and transient ion signals}\label{signals}
\begin{figure}[htbp]
\renewcommand{\baselinestretch}{1}
\begin{center}
\includegraphics[width=8.5cm]{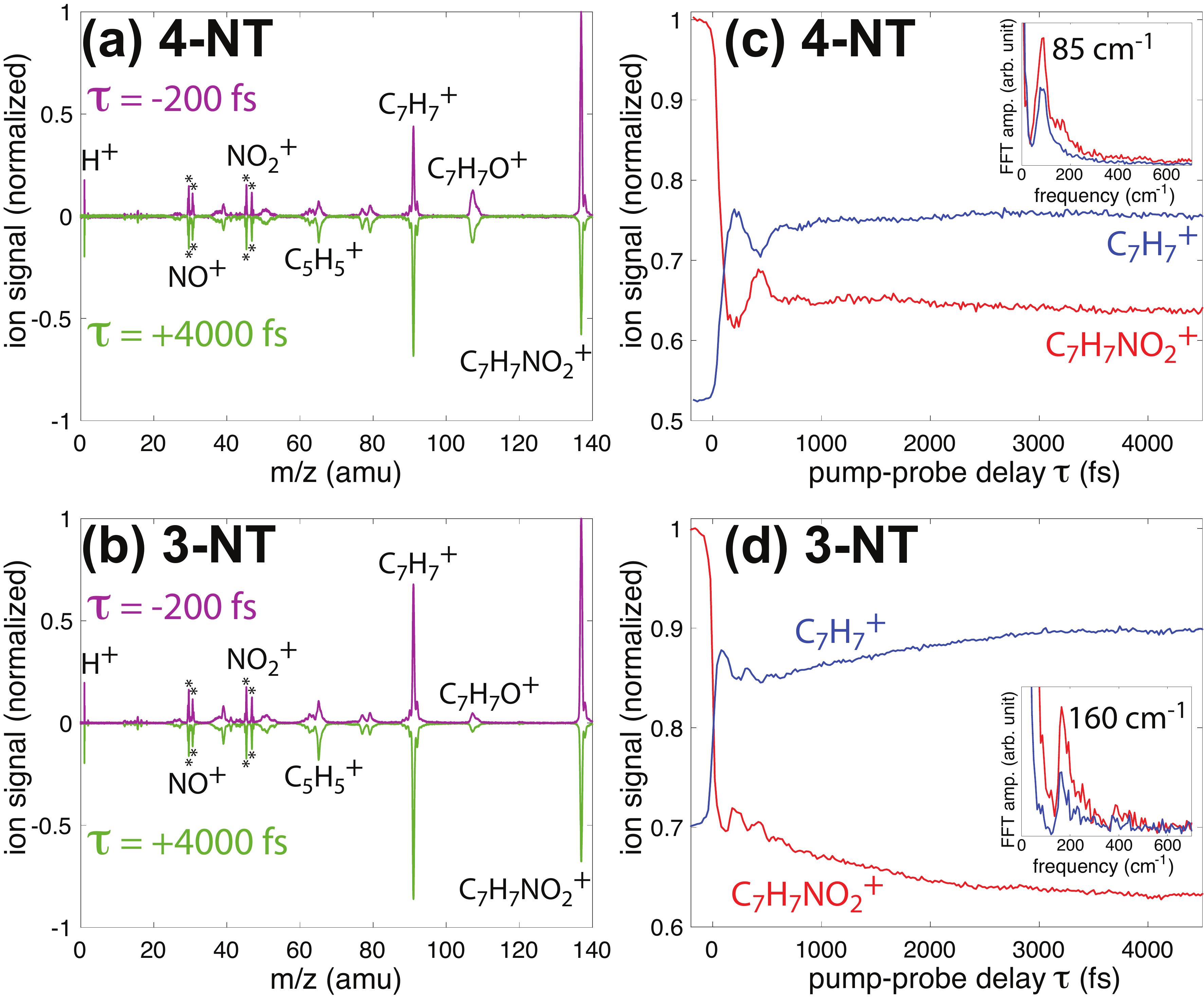}
\end{center}
\caption{\label{transients} Mass spectra of (a) 4-NT and (b) 3-NT taken at pump-probe delays $\tau=-200$ fs in (purple) and $\tau=+4000$ fs (green). Transient ion yields of parent molecular ion (red) and \ce{C7H7+} (blue) in (c) 4-NT and (d) 3-NT as a function of pump-probe delay. Inset: FFT of the transient signals showing the oscillation frequencies.}
\end{figure}

Figures \ref{transients}(a) and (b) display the mass spectra of 4-NT and 3-NT taken with pump intensity $8\times10^{13}$ W cm$^{-2}$ and probe intensity $1\times10^{11}$ W cm$^{-2}$ at pump-probe delays $\tau=-200$ fs (purple) and $\tau=+4000$ fs (green). The mass spectra are normalized to the respective parent ion yields at $\tau=-200$ fs (probe precedes pump). In this situation, all ions are generated solely from the pump because the probe intensity is well below the ionization threshold and the parent molecular ion is the most intense peak for both 3-NT and 4-NT. The predominant formation of parent molecular ion is consistent with previous studies on other molecules under adiabatic ionization conditions.\cite{Lezius2001,Lezius2002,Yatsuhashi2005,Murakami2005,Tanaka2009,Bohinski2013a,Bohinski2013b,Bohinski2014a,Tibbetts2014,Bohinski2014b,Tibbetts2015,Munkerup2017,AmpaduBoateng2018} For both 3-NT and 4-NT at $\tau=+4000$ fs, the parent ion signal is depleted and the \ce{C7H7+} ion signal enhanced, indicating that the weak field probe pulse is capable of exciting ions generated by the pump to form \ce{C7H7+} through cleavage of the \ce{C-NO2} bond. Because the most significant changes in ion yields due to the probe pulse affect the parent ion and \ce{C7H7+}, we will focus on the dynamics of these two ions. Other fragments are visible in the spectra, including \ce{C7H7O+}, formed from the parent ion via nitro-nitrite rearrangement, and \ce{C5H5+}, formed from dissociation of \ce{C7H7+}.\cite{Zhang2012} \ce{NO2+} and \ce{NO+} in both molecules are formed via Columb explosion of a multiply-charged precursor based on the split peaks marked with a $*$ in the mass spectra.\cite{Nibarger2001}

The transient ion signals of the parent ion \ce{C7H7NO2+} (red) and \ce{C7H7+}(blue) as a function of pump-probe delay $\tau$ are shown in Figure \ref{transients}(c) and (d) for 4-NT and 3-NT, respectively. Ion signals in each molecule are normalized to the parent ion yield at $\tau=-200$ fs. While there is a significant depletion of the parent and an increase \ce{C7H7+} at $\tau>0$ in both molecules, the transient dynamics of these species are quite distinct. Out-of-phase oscillations of the parent and \ce{C7H7+} ion signals are visible in each molecule, suggesting (1) that coherent vibrational motions are excited upon ionization of both 4-NT and 3-NT and (2) that \ce{C7H7+} is formed via excitation with the probe pulse to an excited electronic state in the \ce{C7H7NO2+} ion.\cite{Pearson2007,Gonzalez2010,Ho2009,Brogaard2011,Zhu2011,Konar2014,Munkerup2017,Bohinski2014b,Tibbetts2015,AmpaduBoateng2018} Performing a fast Fourier transform (FFT) on the transient signals produced well-resolved peaks at approximately 85 cm$^{-1}$ and 160 cm$^{-1}$ for 4-NT and 3-NT, respectively (insets of Figures \ref{transients} (c) and (d)). The the transient signals remain unchanged at $\tau>1500$ fs for 4-NT and $\tau>+4000$ fs for 3-NT, indicating no further dynamics.

\subsection{Analysis of the oscillatory motions}\label{osc}
To gain further insight into the oscillatory dynamics observed in 4-NT and 3-NT, the transient ion signals for the parent ion and \ce{C7H7+} at $\tau>40$ fs (i.e., after the pump pulse is over) were fit using nonlinear least square methods to the following equations,
 \begin{align}
S_\text{4-NT}(\tau)&=a\exp\left(-\frac{\tau}{T}\right)\left[\sin\left(\frac{2\pi \tau}{t}+\phi\right)+b\right]+c\label{4nt}\\
S_\text{3-NT}(\tau)&=a\exp\left(-\frac{\tau}{T_1}\right)\left[\sin\left(\frac{2\pi \tau}{t}+\phi\right)+b\right]+c+d\exp\left(-\frac{\tau}{T_2}\right)\label{3nt}
\end{align}
where $a$ denotes the oscillation amplitude, $T$ and $T_1$ denote the coherent lifetime in 4-NT (Eq. (\ref{4nt})) and 3-NT (Eq. (\ref{3nt})), $t$ denotes the oscillation period, and $\phi$ denotes the phase. For both molecules, the constant $b$ corresponds to an incoherent contribution to the exponential decay and $c$ corresponds to the final yield as $\tau\to\infty$. The transient signals in 3-NT require a second exponential decay term with amplitude $d$ and lifetime $T_2$ (Eq. (\ref{3nt})) to account for the slow decay until $\tau\sim4000$ fs. 

\begin{figure}[htbp]
\renewcommand{\baselinestretch}{1}
\begin{center}
\includegraphics[width=7cm]{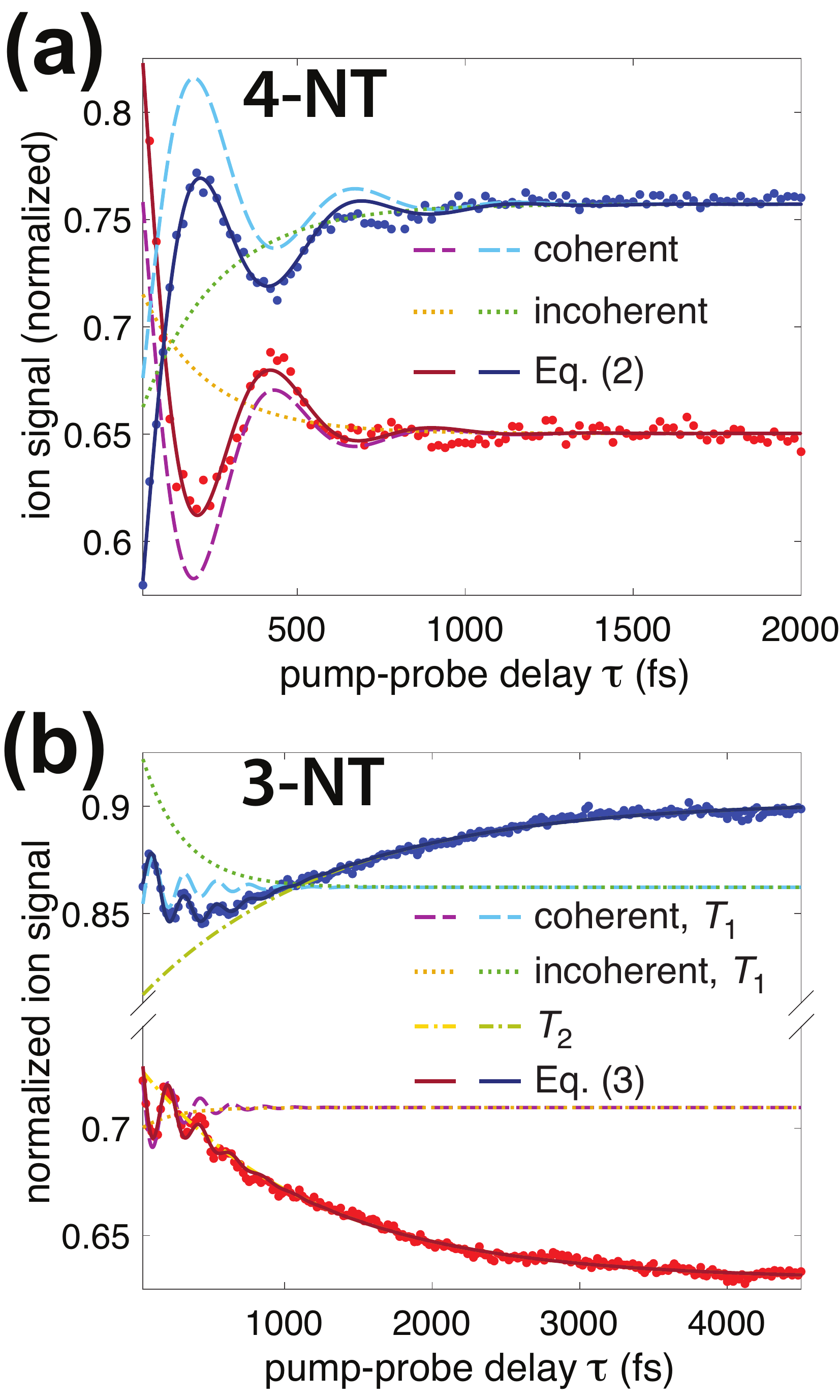}
\end{center}
\caption{\label{fits} Normalized transient ion yields of parent molecular ion (red) and \ce{C7H7+} (blue) with curve fitting components for 4-NT (a) and 3-NT (b), respectively.}
\end{figure}

Figure \ref{fits} shows the fit results for the transient parent and \ce{C7H7+} ion signals in 4-NT (a) and 3-NT (b) from Figure \ref{transients}. Experimental data points are shown as red (parent) and blue (\ce{C7H7+}) dots, and the fit functions to Eqs. (\ref{4nt}) and (\ref{3nt}) for 4-NT and 3-NT as solid lines. The coherent portions of the respective fit functions are shown as magenta and light blue dashed lines for the parent and \ce{C7H7+} ions, and the incoherent portions shown as orange and green dotted lines, respectively. The second exponential contribution in Eq. (\ref{3nt}) for 3-NT is shown as light and dark yellow dash-dot lines for the parent and \ce{C7H7+} ions in Figure \ref{fits}(b). 

To fully characterize the excitation leading to cleavage of the \ce{C-NO2} bond, pump-probe measurements were performed on both molecules at a series of probe intensities from approximately $2\times10^{10}$ to $2\times10^{11}$ W cm$^{-2}$ and fit to Eqs. (\ref{4nt}) or (\ref{3nt}) (Supplemental Information, Figure S3). For both molecules, the following fit parameters were found to be independent of the probe intensity: coherent and incoherent lifetimes, oscillation periods, and phase (summarized in Table \ref{coeff}; all fit parameters to the data in Figure S3 are presented in the Supplemental Information, Tables S1 through S4). The consistent dynamical timescales and phase difference of approximately $\pi$ radians between the parent and \ce{C7H7+} ions indicate that the same excitation processes occur over this range of probe intensities. 
\begin{table}[htbp]
\renewcommand{\baselinestretch}{1}
\begin{tabular}{llrrrrr}
\hline
molecule&species&$T$ (fs)&$T_1$ (fs)&$T_2$ (fs)&$t$ (fs)&$\phi$ (rad)\\
\hline
4-NT& parent&$210\pm10$&&&$480\pm20$&$2.2\pm0.1$\\
& \ce{C7H7+}&$200\pm10$&&&$460\pm10$&$5.2\pm0.1$\\
3-NT&parent&&$220\pm20$&$1100\pm100$&$216\pm3$&$1.4\pm0.2$\\
&\ce{C7H7+}&&$220\pm40$&$1200\pm200$&$220\pm3$&$4.5\pm0.2$\\\hline
\end{tabular}
\caption{\label{coeff} Dynamical timescales and phases obtained by fitting the transient ion signals to Eqs. (\ref{4nt}) and (\ref{3nt}). Errors denote the standard deviation of the fitted coefficient value over the 12 measured probe powers.}
\end{table}
\begin{figure}[htbp]
\renewcommand{\baselinestretch}{1}
\begin{center}
\includegraphics[width=8.5cm]{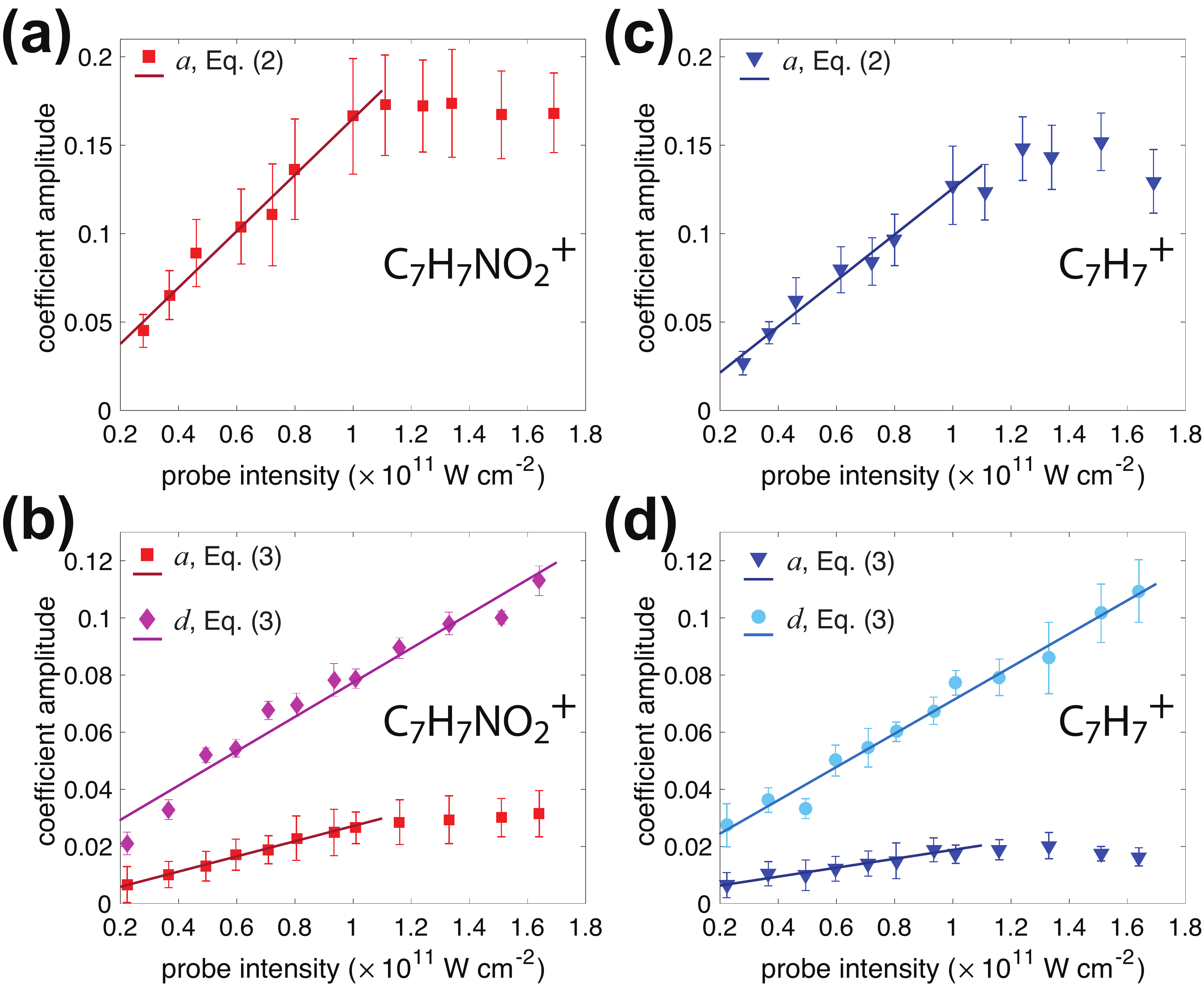}
\end{center}
\caption{\label{coefff} Amplitude coefficients for the parent molecular ion as a function of probe intensity in 4-NT (a) and 3-NT (b). Amplitude coefficients for the \ce{C7H7+} ion as a function of probe intensity in 4-NT (c) and 3-NT (d), respectively. Error bars denote 95$\%$ confidence intervals.}
\end{figure}

The amplitude coefficients corresponding to coherent dynamics ($a$ in Eqs. (\ref{4nt}) and (\ref{3nt})) and slow time decay ($d$ in Eq. (\ref{3nt})) were observed to grow with increasing probe intensity. Figures \ref{coefff}(a) and (b) show the magnitude of the amplitude coefficients for the parent molecular ions of 4-NT and 3-NT, respectively, as a function of probe intensity. The analogous coefficients for the \ce{C7H7+} ions are shown in Figures \ref{coefff}(c) and (d). For all transients, the amplitude coefficients grow linearly with the the probe intensity, as shown by the least squares fit lines. This linear growth indicates a one-photon excitation process, resulting in \ce{C-NO2} bond cleavage. A one-photon excitation was also found to lead to methyl loss in acetophenone radical cation.\cite{Tibbetts2015} While the $a$ coefficients for each transient ion saturate at probe intensities above $\sim10^{11}$ W cm$^{-2}$, the $d$ coefficient in Eq. (\ref{3nt}) for 3-NT continues to grow (magenta and light blue data, Figures \ref{coefff}(b) and (d)). This different behavior in the short- and long-time dynamics of 3-NT suggest that two distinct excitations may contribute to \ce{NO2} loss in 3-NT.

\section{Theoretical results}\label{theoryres}
\begin{figure}[htbp]
\renewcommand{\baselinestretch}{1}
\begin{center}
\includegraphics[width=8.5cm]{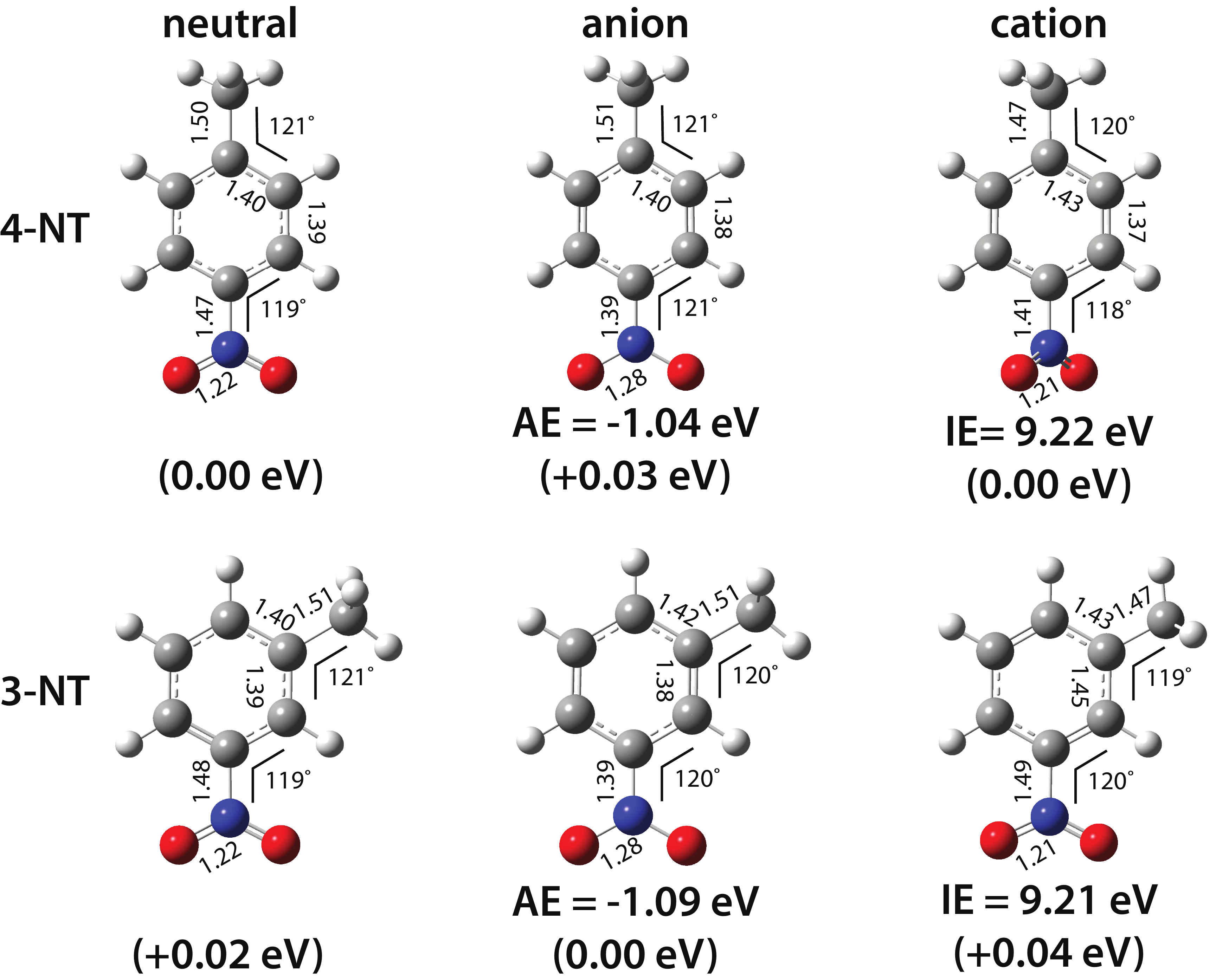}
\end{center}
\caption{\label{structs} Optimized geometrical structures of neutral and singly charged isomers of nitrotoluene. Bond lengths are in {\AA} and angles in degrees. EA: adiabatic electron affinity; IE: adiabatic ionization energy. Numbers in parentheses denote relative energies between isomers in each charge state.}
\end{figure}

Our optimized structures of neutral and charged 3-NT and 4-NT isomers are displayed in Figure \ref{structs}. Structural experimental data have been obtained for 4-NT crystals,\cite{Barve1971} and the measured bond distances agree with our computed values for neutral 4-NT to within $\sim~0.02$ \AA. As can be seen in the figure, the differences in total energy between the two NT isomers are quite small independent of charge, and the largest difference of 0.04 eV belongs to the cation pair. Attachment of an extra electron leads to a significant change in the geometry of the \ce{NO2} group in both 3-NT and 4-NT anions compared to the geometries of their neutral parents as evident in the shortened C$-$N bonds and lengthened N$-$O bonds. Electron attachment also makes 3-NT lower in total energy than 4-NT by 0.03 eV. According to the results of Mulliken analysis, there are 0.6 extra electrons localized over the \ce{NO2} group in both \ce{C7H7NO2-} anions (Supplemental Information, Tables S7 and S8). 

\begin{figure}[htbp]
\renewcommand{\baselinestretch}{1}
\begin{center}
\includegraphics[width=8.5cm]{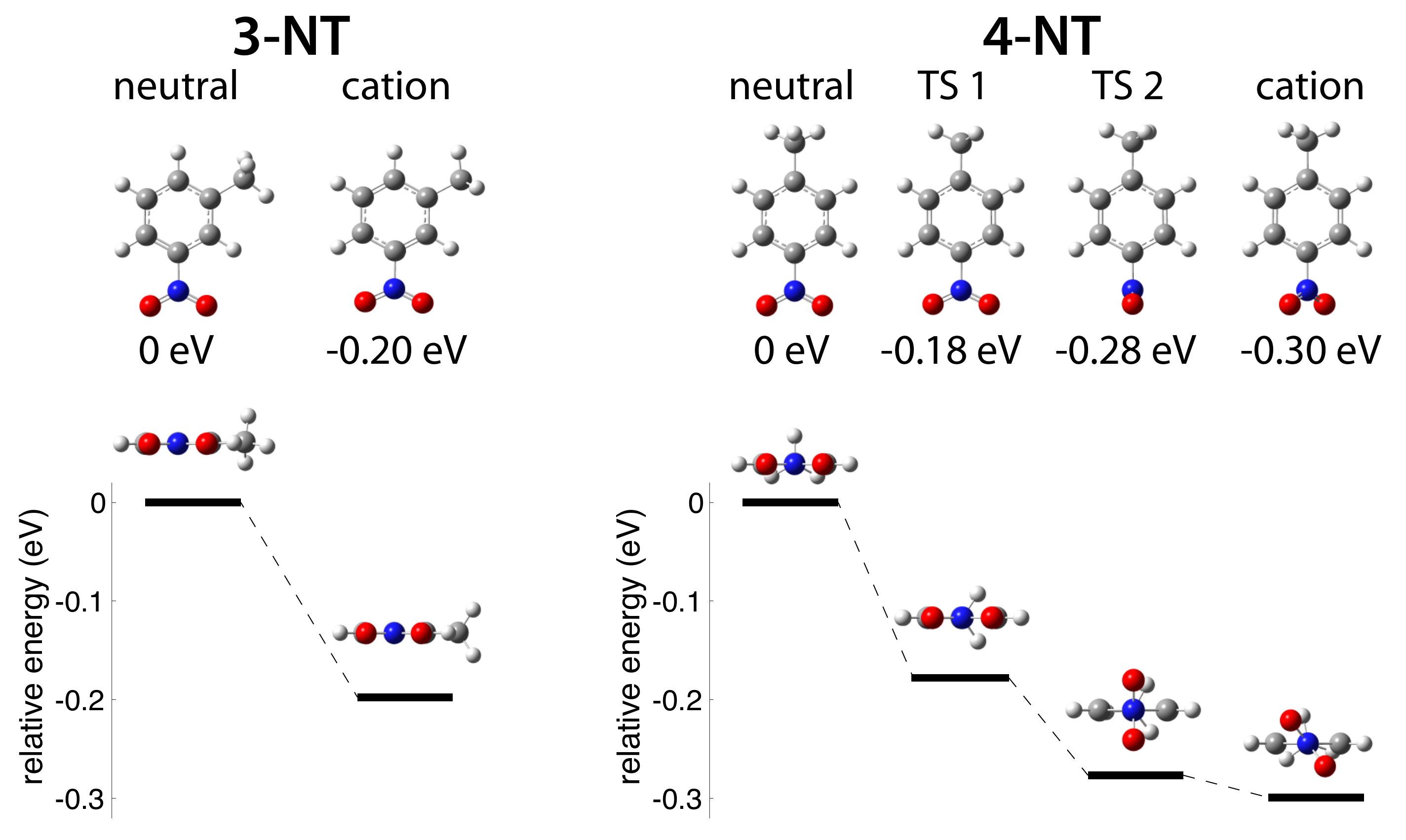}
\end{center}
\caption{\label{relax} The relaxation pathways and energies for the transitions from the neutral geometry to the cation geometries in 3-NT (left) and 4-NT (right).}
\end{figure}

Electron detachment changes the bond lengths in the \ce{C6} rings of both isomers (Figure \ref{structs}) and the ring carries about 0.75 $e$ excessive charge in the 4-NT cation and nearly 0.9 $e$ in the 3-NT cation (Supplemental Information, Tables S7 and S8). Since electron detachment from the neutral 3-NT does not lead to a change in the geometrical topology, there is no energy barrier for a transition from the neutral geometry to the optimal cation geometry. However, this is not the case for the neutral 4-NT, where electron detachment results in the \ce{NO2} plane rotating by $52.5^\circ$ relative to the plane of the phenyl ring. In order to find the pathway from the neutral geometry to the cation geometry, we applied the QST2 approach and found that the pathway proceeds via two transition states as shown in Figure \ref{relax}.

\begin{figure}[htbp]
\renewcommand{\baselinestretch}{1}
\begin{center}
\includegraphics[width=8cm]{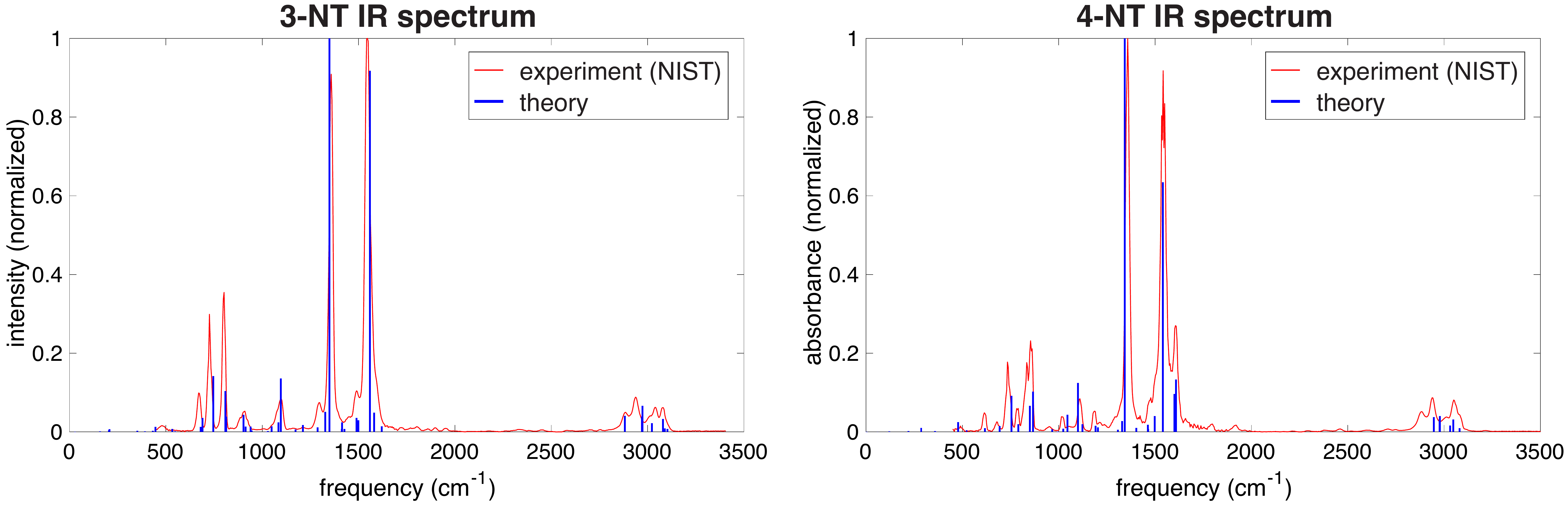}
\end{center}
\caption{\label{IR} Experimental and computed IR spectra for 3-NT (left) and 4-NT (right).}
\end{figure}

Because the observed oscillations in the 4-NT and 3-NT ion yields arise from coherent vibrational motions in the parent radical cations,\cite{Pearson2007,Gonzalez2010,Brogaard2011,Zhu2011,Konar2014,Munkerup2017,Bohinski2014b,Tibbetts2015,AmpaduBoateng2018,Ho2009} it is of interest to determine the vibrational modes in both molecules. The frequencies and intensities of the vibrational modes in both the neutral molecules and their cations were calculated via normal mode analysis. In order to improve comparison with experiments, we have computed third-order anharmonic corrections to the harmonic frequencies of the neutral and cationic 4-NT and 3-NT isomers, whereas the intensities were taken from the harmonic frequency computations. To benchmark the calculated frequencies and intensities, the predicted infrared spectra for the neutral molecules were compared to experimental spectra obtained from NIST\cite{nistspecs} (Figure \ref{IR}). The calculated anharmonic frequencies match the experimental peaks to within 15 cm$^{-1}$ over the frequency range $\sim1000-1600$ cm$^{-1}$ and within 25 cm$^{-1}$ at lower frequencies, indicating the effectiveness of the method and suggesting that the computed cation frequencies should be reasonably accurate. However, adding anharmonic corrections can lead to imaginary (negative) anharmonic frequencies, which is observed for the lowest frequency mode corresponding to the nearly free rotation of the \ce{CH3} group. Full tabulated results of the harmonic and anharmonic vibrational frequencies in the neutral molecules and cations are presented in the Supplemental Information, Tables S9 and S10.

\section{Discussion}\label{disc}
\subsection{Assignment of coherently excited normal modes}
\begin{figure}[htbp]
\renewcommand{\baselinestretch}{1}
\begin{center}
\includegraphics[width=8cm]{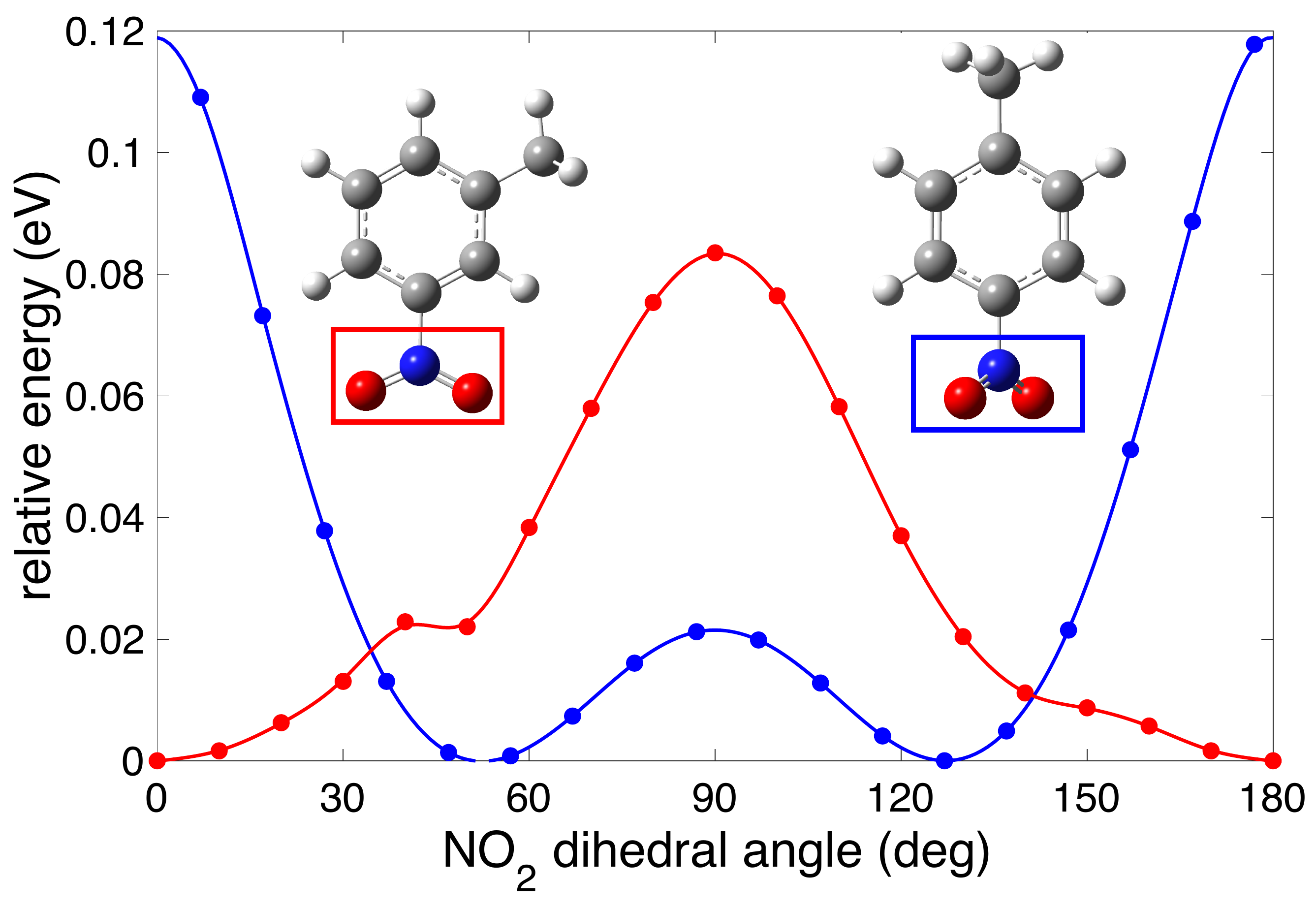}
\end{center}
\caption{\label{PES} Potential energy curves along the \ce{C-C-N-O} dihedral angle for 4-NT (blue) and  3-NT (red).}
\end{figure}

Comparison of the computed relaxation pathways and vibrational frequencies to the observed coherent oscillations in 4-NT and 3-NT radical cations allows for determination of which coherent nuclear motions are excited upon ionization. Based on the experimental observation of an oscillation at 85 cm$^{-1}$ (based on FFT analysis) or $460-480$ fs ($69-73$ cm$^{-1}$, based on curve-fitting) in 4-NT and its $52.5^\circ$ rotation in the \ce{C-C-N-O} torsional angle upon electron detachment (Figure \ref{relax}), it is most likely that the \ce{NO2} torsional mode is excited. The observed frequency is in reasonable agreement with the computed oscillation frequencies in this mode of $59.8$ cm$^{-1}$ in the neutral and $46.1$ cm$^{-1}$ in the ion. To confirm that the \ce{NO2} torsional mode is responsible for the observed coherent oscillations, the potential energy curves along the \ce{NO2} dihedral angle were computed as a function of the \ce{NO2} dihedral angle with steps of 5$^{\circ}$ and 10$^\circ$. Figure \ref{PES} shows the potential energy curves along the \ce{NO2} dihedral angle for both 3-NT (red) and 4-NT (blue). As expected, the potential energy decreases by 0.12 eV in 4-NT as the \ce{NO2} group rotates away from $0^\circ$ to its optimal value at 52.5$^\circ$. The global maximum at $0^\circ$ and local maximum at 90$^\circ$ correspond to TS 1 and TS 2, respectively, of the relaxation pathway in Figure \ref{relax}. It is of interest to note that the shape 4-NT potential energy curve along the \ce{NO2} dihedral angle possesses a remarkable similarity to that for the analogous curve along the \ce{COCH3} dihedral angle in acetophenone, which has a local maximum at $90^\circ$ and global maxima at 0$^\circ$ and 180$^\circ$.\cite{Bohinski2014b,Tibbetts2015} 

\begin{figure}[htbp]
\renewcommand{\baselinestretch}{1}
\begin{center}
\includegraphics[width=8.5cm]{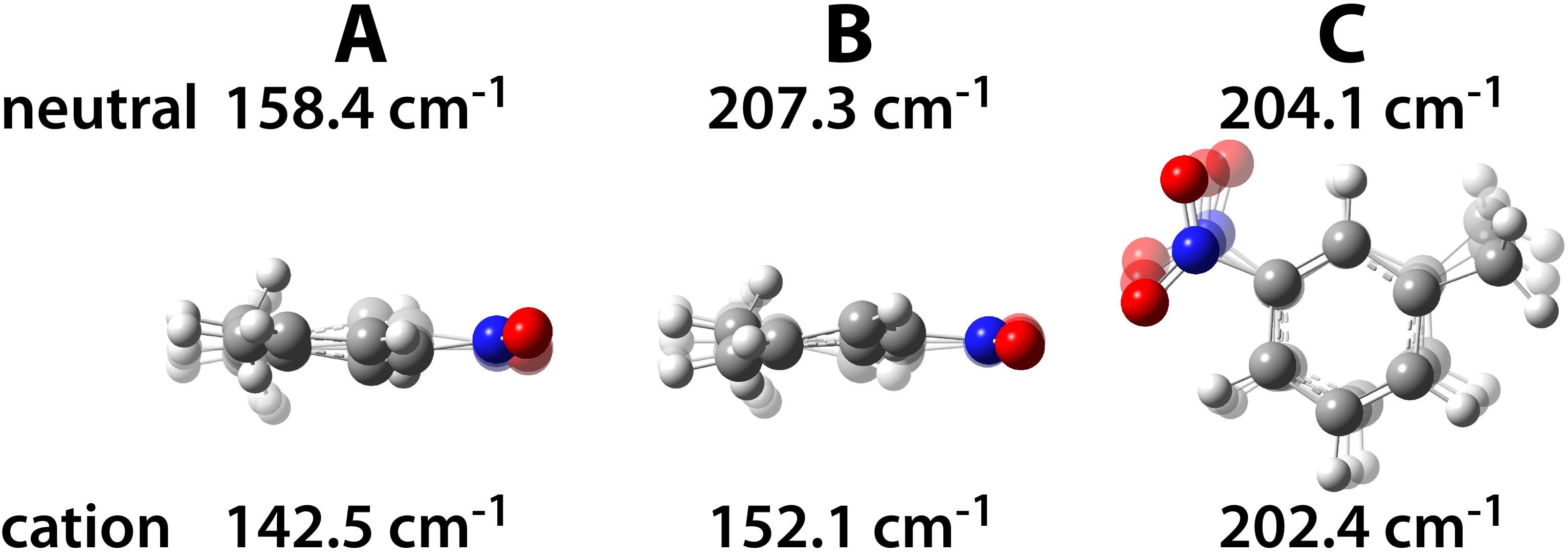}
\end{center}
\caption{\label{mntvibs} Low-frequency bending motions in 3-NT.}
\end{figure}

In contrast to the 4-NT case, the potential energy in the 3-NT radical cation increases as the \ce{NO2} group is rotated away from $0^\circ$ (red curve, Figure \ref{PES}), indicating that the \ce{NO2} torsional mode cannot account for the observed oscillations in 3-NT. This result is consistent with the 160 cm$^{-1}$ oscillations observed in the 3-NT ion yields because the \ce{NO2} torsional mode would be expected at 29 cm$^{-1}$ in the neutral and 40 cm$^{-1}$ in the cation according to our computational results (Table S10). Instead, we consider a group of three normal modes with computed frequencies in the range of $158-207$ cm$^{-1}$ in the neutral and $143-202$ cm$^{-1}$ in the cation (Table S10) to account for the 160 cm$^{-1}$ oscillations. These modes correspond to the low-frequency bending motions shown in Figure \ref{mntvibs}. Because modes \textbf{A} and \textbf{B} correspond to out-of-plane bending motions in the benzene ring, neither is likely to be excited in our experiments because the benzene ring does not change from its planar geometry upon ionization (Figures \ref{structs} and \ref{relax}). Thus, we suggest that mode \textbf{C} corresponding to the in-plane bending motion of the \ce{NO2} and \ce{CH3} moieties gives rise to the observed oscillations. This mode assignment is supported by the changes in bond lengths and angles involving the benzene ring, \ce{NO2}, and \ce{CH3} groups in 3-NT when going from the neutral to cation geometry (Figure \ref{structs}).

The coherent excitation of a torsional motion in ionized 4-NT may be expected because coherent torsional mode excitation has been observed in a number of molecules including acetophenone and its derivatives,\cite{Bohinski2014b,Tibbetts2015, Zhu2011, Konar2014} 1,3-dibromopropane,\cite{Brogaard2011} and azobenzene. \cite{Ho2009, Munkerup2017} The lack of torsional mode excitation in 3-NT upon ionization also resembles the case of 3-methylacetophenone, where no coherent oscillations were observed.\cite{Konar2014} The authors attributed the lack of oscillations in 3-methylacetophenone to the increase in the potential energy upon rotation of the acetyl group away from the planar geometry, which is similar to the potential energy curve in 3-NT (Figure \ref{PES}). Unlike the latter results, we do see coherent oscillations in the 3-NT ion yields from excitation of the in-plane bending mode \textbf{C} shown in Figure \ref{mntvibs}. We attribute this ability to resolve such small-amplitude oscillations to the use of a 1500 nm pump wavelength that ensures adiabatic ionization and predominant population of the ground state molecular ion.\cite{Lezius2001,Lezius2002,Bohinski2014b,AmpaduBoateng2018}

\subsection{Dynamical timescales in 3-NT and 4-NT radical cations}
\begin{figure}[htbp]
\renewcommand{\baselinestretch}{1}
\begin{center}
\includegraphics[width=8.5cm]{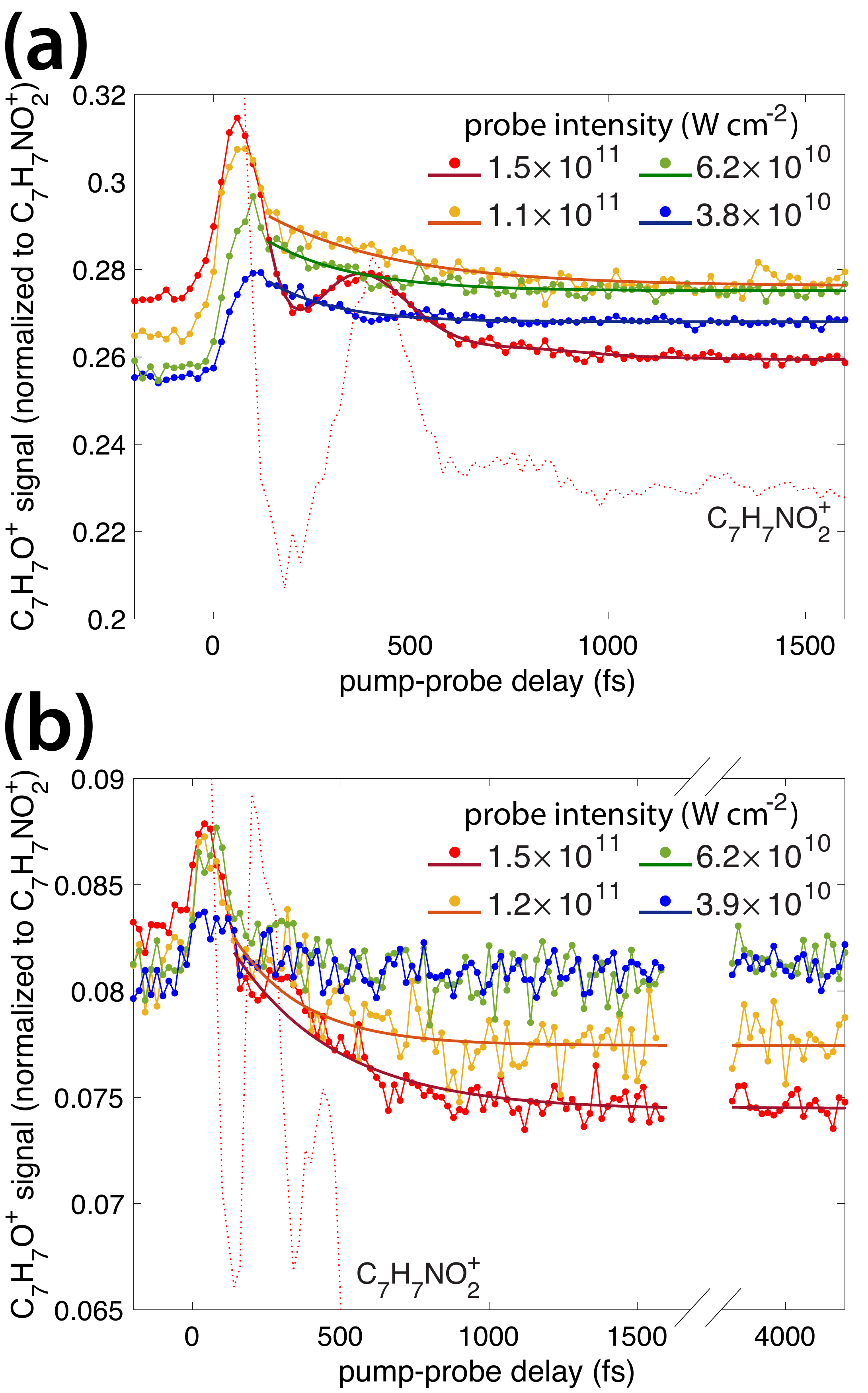}
\end{center}
\caption{\label{nnr} Transient ion signals of \ce{C7H7O+} in (a) 4-NT and (b) 3-NT at selected probe intensities (dots), indicated by color in the legend. Fit functions to Eq. (\ref{3nt}) or a decaying exponential are shown as solid lines. The signals in 3-NT at intensities below $10^11$ W/cm$^2$ were too noisy for curve fitting. The transient signal of the parent molecular ion is shown as the dotted line.}
\end{figure}

The oscillations in both 3-NT and 4-NT decay with similar time constants of 220 fs and 200 fs, respectively. This short coherence lifetime stands in contrast to the longer coherent lifetimes of torsional wavepackets in acetophenone ($560-600$ fs)\cite{Bohinski2014b,Tibbetts2015} and azobenzene ($880-1000$ fs).\cite{Munkerup2017} Unlike the latter molecules, which are not known to undergo rearrangement reactions, the nitro group can undergo the nitro-nitrite rearrangement (NNR) reaction (\ce{NO2 -> ONO}). This rearrangement would change the topology of the molecule and thus be expected to destroy the initially excited coherent nuclear motion. The observation of \ce{C7H7O+} arising from \ce{NO} loss following NNR in our mass spectra (Figure \ref{transients}) indicates that NNR takes place in the 4-NT and 3-NT radical cations, so this rearrangement may be expected to cause the faster decoherence as compared to other aromatic molecules. However, the \ce{C7H7O+} transient dynamics (Figure \ref{nnr}) suggest that NNR does not drive the fast decoherence in either 4-NT or 3-NT. If the NNR reaction were the primary cause of decoherence, an exponentially increasing yield of \ce{C7H7O+} with a similar time constant of $\sim200$ fs as the wavepacket decay time constant would be expected. 

Instead, the \ce{C7H7O+} transients in both 4-NT and 3-NT exhibit completely different dynamics from the respective parent molecular ions and \ce{C7H7+} fragments. While parent and \ce{C7H7+} dynamics are the same at all probe intensities in a given isomer (Supporting Information, Figure S3), the \ce{C7H7O+} dynamics in both isomers are sensitive to the probe intensity. At intensities above $6\times10^{10}$ W cm$^{-2}$, both 4-NT and 3-NT produce a spike in \ce{C7H7O+} yield approximately $60-80$ fs after the ionization event, followed by exponential decay of the signal with time constants ranging from $170-350$ fs (solid lines in Figure \ref{nnr}; fit coefficients given in the Supporting Information, Tables S11 and S12). This result suggests that excitation of both 4-NT and 3-NT radical cations at short time-delays can facilitate the NNR reaction, and that the excitation probability quickly decreases at longer time-delays. In 4-NT, the \ce{C7H7O+} transient exhibits similar oscillatory dynamics in-phase with the parent molecular ion at a high probe intensity of $1.5 \times10^{11}$ W cm$^{-2}$, as seen in the fit to Eq. (\ref{3nt}) (solid red line), indicating that the NNR reaction can also take place on the ground electronic state of the 4-NT cation. However, the lack of these dynamics at lower probe intensities suggest that spontaneous rearrangement on the ground state is not the primary NNR pathway in 4-NT. In all cases for both isomers, the observed \ce{C7H7O+} dynamics do not suggest that NNR occurs spontaneously within 200 fs to cause wavepacket decoherence.

 The complex dynamics of the \ce{C7H7O+} ion in 4-NT and 3-NT suggest that multiple pathways involving NNR exist in the respective radical cations, separate from the coherent excitation pathways leading to \ce{C-NO2} bond cleavage. The distinct \ce{C7H7O+} dynamics in each isomer further highlight the different dynamics of 4-NT and 3-NT apparent in the separate coherent vibrations excited in the respective radical cations. Furthermore, the incoherent 1.1 ps decay of the parent ion yield and increase of the \ce{C7H7+} yield in 3-NT suggests that an additional dynamical relaxation process facilitates the excitation leading to \ce{C-NO2} bond cleavage in 3-NT radical cation, while the analogous process is absent in 4-NT. The observation that the amplitude coefficient associated with the slow decay in 3-NT continues to grow at high probe powers where the amplitude coefficient associated with the coherent excitation is saturated suggests that two different excitation processes, possibly involving distinct excited states, may form \ce{C7H7+} in 3-NT. Determination of all of the pathways leading to both NNR and \ce{C-NO2} cleavage will require high-level quantum chemical calculations of both the ground and excited state potential energy surfaces along the relevant reaction coordinates. We plan to carry out these and other calculations in order to gain a greater understanding of the excitation and dissociation mechanisms involved.

\section{Conclusions}\label{con}
The ultrafast dynamics of 3- and 4-nitrotoluene radical cations was investigated with femtosecond pump-probe measurements and high-level DFT calculations. Oscillations in the parent and \ce{C7H7+} ion yields with pump-probe delay arising from coherent vibrational excitations were present in both molecules, with similar coherent lifetimes of approximately 200 fs. The distinct oscillation periods of 470 fs and 216 fs in 4-NT and 3-NT, respectively, were attributed to excitation of the \ce{NO2} torsional mode in 4-NT and an in-plane bending mode involving the \ce{NO2} and \ce{CH3} moieties in 3-NT. These normal mode assignments were supported by a series of DFT calculations at the B3LYP/Def2-TZVPP level of the ionization potentials, relaxation pathways, and normal mode frequencies. Loss of \ce{NO2} from the parent ions of both 3-NT and 4-NT to form \ce{C7H7+} was found to arise from a one-photon excitation of the initially formed ground state molecular ion based on the linear growth of the fitted amplitude coefficients with the probe pulse intensity. These results show that coherent nuclear dynamics contributes to \ce{C-NO2} homolysis in both nitrotoluene radical cations and open up the potential for further investigation of coherent control schemes to manipulate dissociation pathways in nitroaromatic and other energetic molecules.

{\bf Supplementary Material} Pulse characterization; tabulated data of pump-probe nonlinear least squares coefficients; tabulated computational results of molecular geometries, charge states, and vibrational frequencies.

{\bf Acknowledgements} The authors acknowledge support from the Army Research Office through Contract W911NF-18-1-0051 and from Virginia Commonwealth University. P. J. acknowledges support from the U.S. Department of Energy, Office of Basic Energy Sciences, Division of Materials Sciences and Engineering under Award No. DE-FG02-96ER45579.

\bibliography{explosives_bib.bib}

\end{document}